\documentclass[doublecol]{epl2}
\usepackage{pdflscape}
\usepackage{amsmath}
\usepackage{amssymb}

\newcommand{\CC}{\Lambda}

\newcommand{\nueff}{\nu_{\rm eff}}

\newcommand{\be}{\begin{equation}}
\newcommand{\ee}{\end{equation}}
\def\beq{\begin{equation}}
\def\eeq{\end{equation}}
\def\ber{\begin{eqnarray}}
\def\eer{\end{eqnarray}}

\newcommand{\rv}{\rho_{\rm vac}}
\newcommand{\rvo}{\rho^0_{\rm vac}}
\newcommand{\mPl}{m_{\rm Pl}}

\newcommand{\astar}{a_{*}}
\newcommand{\zstar}{z_{*}}

\title{Running vacuum against the $H_0$ and $\sigma_8$ tensions}

\author{Joan Sol\`a Peracaula$^{a}$,  Adri\`a G\'omez-Valent$^{b}$, Javier de Cruz P\'erez$^{a}$ and Cristian Moreno-Pulido$^{a}$} 
\shortauthor{J. Sol\`a Peracaula \etal}

\institute{$^{a}$ Departament de F\'isica Qu\`antica i Astrof\'isica, and Institute of Cosmos Sciences, Universitat de Barcelona, \\ Av. Diagonal 647, 
$^{b}$ Institut f\"{u}r Theoretische Physik, Ruprecht-Karls-Universit\"{a}t Heidelberg, Philosophenweg 16, D-69120 Heidelberg, Germany
}


\abstract{The cosmological term, $\Lambda$, was introduced $104$ years ago by Einstein in his gravitational field equations. Whether  $\Lambda$  is a rigid quantity or a dynamical variable in cosmology has been a matter of debate for many years, especially after the introduction of the general notion of dark energy (DE).  $\Lambda$ is associated to the vacuum energy density, $\rho_{\rm vac}$, and one may expect that it evolves slowly with the cosmological expansion. Herein we present a devoted study testing this possibility using the promising class of running vacuum models (RVM's).  We use a large string SNIa+BAO+$H(z)$+LSS+CMB of modern cosmological data, in which for the first time the CMB part involves the full Planck 2018 likelihood for these models.  We test the dependence of the results on the threshold  redshift $\zstar$  at which the vacuum dynamics is activated in the recent past and  find positive signals up to  $\sim4.0\sigma$ for  $\zstar\simeq 1$. The  RVM's prove very competitive against the standard  $\Lambda$CDM model and give a handle for solving  the $\sigma_8$  tension and alleviating the  $H_0$ one. }

\begin{document}

\maketitle

\section{Introduction}
\label{intro}

Despite Einstein's original formulation\,\cite{Einstein1917}, in which the cosmological term $\CC$  is treated as a strict constant in the gravitational field equations, the idea that $\CC$ (and its associated vacuum energy density $\rv$)  can be a  dynamical quantity should be most natural in the context of an expanding universe.  This point of view has led to the notion of dynamical dark energy (DDE) in its multifarious forms\,\cite{CCP,DEBook}.  Herein, however, we stick to the notion of dynamical vacuum energy  (DVE) as the ultimate cause of DDE.  Despite the fact that $\rv$ has long been associated with the so-called cosmological constant problem\,\cite{Weinberg,JSPRev2013}, which involves severe fine-tuning of the parameters, such a conundrum actually underlies all of the DE models known up to date, with no exception\,\cite{JSPRev2013}.  In addition, recent calculations of $\rv$ in the context of quantum field theory (QFT) in curved spacetime have brought new light into this problem\, \cite{Cristian2020} and suggest that if  the vacuum energy density (VED) is renormalized using an appropriate regularization procedure, it evolves in a mild way as a series of powers of the Hubble rate $H$ and its cosmic time derivatives: $\rv(H,\dot{H},...)$, denoted $\rv(H)$ for short.  This fact was long foreseen from general renormalization group arguments which led to the notion of running vacuum models (RVM's), see the reviews \cite{JSPRev2013,SolGom2015} and references therein.
\renewcommand{\arraystretch}{0.7}
\begin{table*}[t!]
\begin{center}
\resizebox{1\textwidth}{!}{
\begin{tabular}{|c  |c | c |  c | c | c  |c  |}
 \multicolumn{1}{c}{} & \multicolumn{4}{c}{Baseline}
\\\hline
{\scriptsize Parameter} & {\scriptsize GR-$\Lambda$CDM}  & {\scriptsize type I RRVM} & {\scriptsize type I RRVM$_{\rm thr.}$}  &  {\scriptsize type II RRVM} &  {\scriptsize BD-$\Lambda$CDM}
\\\hline
{\scriptsize $H_0$(km/s/Mpc)}  & {\scriptsize $68.37^{+0.38}_{-0.41}$} & {\scriptsize $68.17^{+0.50}_{-0.48}$} & {\scriptsize $67.63^{+0.42}_{-0.43}$}  & {\scriptsize $69.02^{+1.16}_{-1.21}$} & {\scriptsize $69.30^{+1.38}_{-1.33}$}
\\\hline
{\scriptsize$\omega_b$} & {\scriptsize $0.02230^{+0.00019}_{-0.00018}$}  & {\scriptsize $0.02239^{+0.00023}_{-0.00024}$} & {\scriptsize $0.02231^{+0.00020}_{-0.00019}$}  &  {\scriptsize $0.02245^{+0.00025}_{-0.00027}$} & {\scriptsize $0.02248\pm 0.00025$}
\\\hline
{\scriptsize$\omega_{dm}$} & {\scriptsize $0.11725^{+0.00094}_{-0.00084}$}  & {\scriptsize $0.11731^{+0.00092}_{-0.00087}$} & {\scriptsize $0.12461^{+0.00201}_{-0.00210}$}  &  {\scriptsize $0.11653^{+0.00158}_{-0.00160}$} &   {\scriptsize $0.11629^{+0.00148}_{-0.00151}$}
\\\hline
{\scriptsize$\nu_{\rm eff}$} & {-}  & {\scriptsize $0.00024^{+0.00039}_{-0.00040}$} & {\scriptsize $0.02369^{+0.00625}_{-0.00563}$}  &  {\scriptsize $0.00029\pm 0.00047$} & {-}
\\\hline
{\scriptsize$\epsilon_{\rm BD}$} & {-}  & {-} & {-}  &  {-} & {\scriptsize $-0.00109\pm ^{+0.00135}_{-0.00141}$}
\\\hline
{\scriptsize$\varphi_{\rm ini}$} & {-}  & {-} & {-}  &  {\scriptsize $0.980^{+0.031}_{-0.027}$} & {\scriptsize $0.972^{+0.030}_{-0.037}$}
\\\hline
{\scriptsize$\varphi_0$} & {-}  & {-} & {-}  &  {\scriptsize $0.973^{+0.036}_{-0.033}$} & {\scriptsize $0.963^{+0.036}_{-0.041}$}
\\\hline
{\scriptsize$\tau_{\rm reio}$} & {{\scriptsize$0.049^{+0.008}_{-0.007}$}} & {{\scriptsize$0.051^{+0.008}_{-0.009}$}} & {{\scriptsize$0.058^{+0.007}_{-0.009}$}}  &   {{\scriptsize$0.051\pm 0.008$}} & {{\scriptsize$0.051\pm 0.008$}}
\\\hline
{\scriptsize$n_s$} & {{\scriptsize$0.9698^{+0.0039}_{-0.0036}$}}  & {{\scriptsize$0.9716^{+0.0044}_{-0.0047}$}} & {{\scriptsize$0.9703\pm 0.038$}} &   {{\scriptsize$0.9762^{+0.0081}_{-0.0091}$}} &
\\\hline
{\scriptsize$\sigma_8$}  & {{\scriptsize$0.796\pm 0.007$}}  & {{\scriptsize$0.789^{+0.013}_{-0.014}$}} & {{\scriptsize$0.768^{+0.010}_{-0.009}$}}  &   {{\scriptsize$0.791^{+0.013}_{-0.012}$}} & {{\scriptsize$0.790^{+0.013}_{-0.012}$}}
\\\hline
{\scriptsize$S_8$}  & {{\scriptsize$0.796\pm 0.011$}}  & {{\scriptsize$0.791^{+0.014}_{-0.013}$}} & {{\scriptsize$0.797^{+0.012}_{-0.011}$}}  &   {{\scriptsize$0.781^{+0.021}_{-0.020}$}} & {{\scriptsize$0.777^{+0.021}_{-0.022}$}}
\\\hline
{\scriptsize$r_s$ (Mpc)}  & {{\scriptsize$147.90^{+0.30}_{-0.31}$}}  & {{\scriptsize$147.99^{+0.35}_{-0.36}$}} & {{\scriptsize$147.81\pm 0.30$}}  &   {{\scriptsize$146.30^{+2.39}_{-2.30}$}} & {{\scriptsize$145.72^{+2.44}_{-2.90}$}}
\\\hline
{\scriptsize$\chi^2_{\rm min}$}  & {{\scriptsize 2290.20}}  & {{\scriptsize 2289.72}} & {{\scriptsize 2272.44}}  &   {{\scriptsize 2288.74}} & {{\scriptsize 2289.40}}
\\\hline
{\scriptsize$\Delta{\rm DIC}$}  & {-}  & {{\scriptsize -2.70}} & {{\scriptsize +13.82}}  &   {{\scriptsize -4.59}} & {{\scriptsize -3.53}}
\\\hline
\end{tabular}}
\end{center}
\caption{The mean values and 68.3\% confidence limits for the models under study using our Baseline dataset, which is almost the same as the one employed in \cite{BD-RVM}, with few changes: (i) for the eBOSS survey we have replaced the data from \cite{MarineBOSS} with the one from \cite{NeveuxeBOSS}; (ii) the LyF data have been updated, replacing \cite{AgatheLyF} with \cite{BourbouxLyF}; (iii) finally, we have replaced the two $f\sigma_8$ data points \cite{Qinfs8,Shifs8} with the one provided in \cite{Saidfs8}. We display the fitting values for the usual parameters, to wit: $H_0$, the reduced density parameter for baryons ($w_b = \Omega^0_b{h^2}$) and CDM ($w_{dm} = \Omega^0_{dm}{h^2}$), with $\Omega_i^0=8\pi G_N\rho^0_i/3H_0^2$ and $h$ the reduced Hubble constant, the reionization optical depth $\tau_{\rm reio}$, the spectral index $n_s$ and the current matter density rms fluctuations within spheres of radius $8h^{-1}$~Mpc, i.e.  $\sigma_8$. We include also a couple of useful derived parameters, namely: the sound horizon at the baryon drag epoch $r_s$ and $S_8\equiv \sigma_8\sqrt{\Omega^0_m/0.3}$. For all the RRVM's we show $\nu_{\rm eff}$, and for the type II {and BD-$\Lambda$CDM}\cite{BD-RVM} we also report the initial and current values of $\varphi$, $\varphi_{\rm ini}$ and $\varphi_0$, respectively. {The parameter $\epsilon_{\rm BD}\equiv1/\omega_{\rm BD}$ (inverse of the Brans-Dicke parameter\,\cite{BD}) controls the dynamics of the scalar field\cite{BD-RVM}.}
We provide the corresponding values of $\chi^2_{\rm min}$ and $\Delta$DIC.}
\label{tableFit1}
\end{table*}
%
 For the current universe, the leading VED term is constant but the next-to-leading one is dynamical, specifically it evolves as a power  $\sim H^2$ with a small coefficient $|\nu|\ll1$. For the early universe, terms of order $\sim H^4$ or higher appear and these can trigger inflation\,\cite{LiBasSol2013,SolGom2015,Sola2015}.  It is remarkable that the fourth power $H^4$ can be motivated within  the context of string theory calculations at low energy (meaning near the Planck scale)\,\cite{BasMavroSola2019}, what reveals  a distinctive mechanism of inflation different from that of Starobinsky inflation\,\cite{staro}, for example.   See\,\cite{Cristian2020,NickJoan2020} for a detailed discussion. Here, however, we will concentrate on the post-inflationary universe, where only the leading power $\sim H^2$ is involved in the dynamics of  $\rv$. A variety of phenomenological analyses have supported this possibility in recent years
 \,\cite{rvmfit1,rvmPerturb,rvmfit2,GomSolBas2015,BD-RVM,PericoTamayo2017,GengLeeYin2017}.

 In this Letter, we present a devoted study of the class of RVM's based on a large and updated string SNIa+BAO+$H(z)$+LSS+CMB of modern cosmological observations, in which for the first time the CMB part involves the full Planck 2018 likelihood. We also test the potential dependence of the results on the threshold redshift $\zstar$  at which the DVE becomes  activated in the recent past.  We find that different  RVM's prove  very helpful to alleviate the persisting tensions between the concordance $\CC$CDM model and the structure formation data (the so-called $\sigma_8$ tension) and the mismatch between the local values of the Hubble parameter and those derived from the  CMB\,\cite{Planck2018} (the $H_0$ tension).  These tensions are well described in the literature, see e.g. the reviews\,\cite{tensions,Intertwined}.  Many models in the market try to address them, see e.g. Ref.\,\cite{tensions2} and the long list of references therein.

 In the current (fully updated) study we find significant signals of DVE (using  $\zstar\simeq 1$) at $\sim 3.6\sigma$ c.l., which can be enhanced up to  $\sim4.0\sigma$.  Finally, we show that the  RVM's provide an overall fit to the cosmological data which is comparable or significantly better than in the $\CC$CDM case, as confirmed by calculating the relative  Deviance Information Criterion (DIC) differences obtained form the Monte Carlo chains of our numerical analysis.

\section{Running vacuum Universe}

As indicated, the total vacuum part of the energy-momentum tensor,  $T_{\mu \nu}^{\rm vac}$,  can be appropriately  renormalized into a finite quantity which depends on the Hubble rate $H$ and its time derivatives\, \cite{Cristian2020}.  The corresponding  $00$-component  defines the vacuum energy density (VED),  $\rv(H)$.  Let us denote by  $\rvo\equiv\rv(H_0)=\CC/(8\pi G_N)$ ($G_N$ being Newton's constant)  the current value of the latter, with $H_0$  today's value of the Hubble parameter and $\CC$ the measured  cosmological constant term.  We define two types of DVE scenarios.  In type I scenario the vacuum is in interaction with matter, whereas in type II  matter is conserved at the expense of an exchange between the vacuum and a slowly evolving gravitational coupling $G (H)$.   The combined cosmological `running' of these quantities  insures the accomplishment of the Bianchi identity (and the  local conservation law).


\renewcommand{\arraystretch}{0.65}
\begin{table*}[t!]
\begin{center}
\resizebox{1\textwidth}{!}{
\begin{tabular}{|c  |c | c |  c | c | c  |c  |}
 \multicolumn{1}{c}{} & \multicolumn{4}{c}{Baseline + $H_0$}
\\\hline
{\scriptsize Parameter} & {\scriptsize GR-$\Lambda$CDM}  & {\scriptsize type I RRVM} & {\scriptsize type I RRVM$_{\rm thr.}$}  &  {\scriptsize type II RRVM} &  {\scriptsize BD-$\Lambda$CDM}
\\\hline
{\scriptsize $H_0$ (km/s/Mpc)}  & {\scriptsize $68.75^{+0.41}_{-0.36}$} & {\scriptsize $68.77^{+0.49}_{-0.48}$} & {\scriptsize $68.14^{+0.43}_{-0.41}$}  & {\scriptsize $70.93^{+0.93}_{-0.87}$}  & {\scriptsize $71.23^{+1.01}_{-1.02}$}
\\\hline
{\scriptsize$\omega_b$} & {\scriptsize $0.02240^{+0.00019}_{-0.00021}$}  & {\scriptsize $0.02238^{+0.00021}_{-0.00023}$} & {\scriptsize $0.02243^{+0.00019}_{-0.00018}$}  &  {\scriptsize $0.02269^{+0.00025}_{-0.00024} $}  & {\scriptsize $0.02267^{+0.00026}_{-0.00023} $}
\\\hline
{\scriptsize$\omega_{dm}$} & {\scriptsize $0.11658^{+0.00080}_{-0.00083}$}  & {\scriptsize $0.11661^{+0.00084}_{-0.00085}$} & {\scriptsize $0.12299^{+0.00197}_{-0.00203}$}  &  {\scriptsize $0.11602^{+0.00162}_{-0.00163}$}  & {\scriptsize $0.11601^{+0.00161}_{-0.00157}$}
\\\hline
{\scriptsize$\nu_{\rm eff}$} & {-}  & {\scriptsize $-0.00005^{+0.00040}_{-0.00038}$} & {\scriptsize $0.02089^{+0.00553}_{-0.00593}$}  &  {\scriptsize $0.00038^{+0.00041}_{-0.00044}$}  & {-}
\\\hline
{\scriptsize$\epsilon_{\rm BD}$} & {-}  & {-} & {-}  &  {-} & {\scriptsize $-0.00130\pm ^{+0.00136}_{-0.00140}$}
\\\hline
{\scriptsize$\varphi_{\rm ini}$} & {-}  & {-} & {-}  &  {\scriptsize $0.938^{+0.018}_{-0.024}$}  & {\scriptsize $0.928^{+0.024}_{-0.026}$}
\\\hline
{\scriptsize$\varphi_0$} & {-}  & {-} & {-}  &  {\scriptsize $0.930^{+0.022}_{-0.029}$}  & {\scriptsize $0.919^{+0.028}_{-0.033}$}
\\\hline
{\scriptsize$\tau_{\rm reio}$} & {{\scriptsize$0.050^{+0.008}_{-0.007}$}} & {{\scriptsize$0.049^{+0.009}_{-0.008}$}} & {{\scriptsize$0.058^{+0.008}_{-0.009}$}}  &   {{\scriptsize$0.052\pm 0.008$}}  & {{\scriptsize$0.052\pm 0.008$}}
\\\hline
{\scriptsize$n_s$} & {{\scriptsize$0.9718^{+0.0035}_{-0.0038}$}}  & {{\scriptsize$0.9714\pm 0.0046$}} & {{\scriptsize$0.9723^{+0.0040}_{-0.0039}$}} &   {{\scriptsize$0.9868^{+0.0072}_{-0.0074}$}}  & {{\scriptsize$0.9859^{+0.0073}_{-0.0072}$}}
\\\hline
{\scriptsize$\sigma_8$}  & {{\scriptsize$0.794\pm 0.007$}}  & {{\scriptsize$0.795\pm 0.013$}} & {{\scriptsize$0.770\pm 0.010$}}  &   {{\scriptsize$0.794^{+0.013}_{-0.012}$}}  & {{\scriptsize$0.792^{+0.013}_{-0.012}$}}
\\\hline
{\scriptsize$S_8$}  & {{\scriptsize$0.788^{+0.010}_{-0.011}$}}  & {{\scriptsize$0.789\pm 0.013$}} & {{\scriptsize$0.789\pm 0.011$}}  &   {{\scriptsize$0.761^{+0.018}_{-0.017}$}}  & {{\scriptsize$0.758^{+0.019}_{-0.018}$}}
\\\hline
{\scriptsize$r_s$ (Mpc)}  & {{\scriptsize$147.97^{+0.29}_{-0.31}$}}  & {{\scriptsize$147.94^{+0.35}_{-0.36}$}} & {{\scriptsize$147.88^{+0.33}_{-0.29}$}}  &   {{\scriptsize$143.00^{+1.54}_{-1.96}$}}  & {{\scriptsize$142.24^{+1.99}_{-2.12}$}}
\\\hline
{\scriptsize$\chi^2_{\rm min}$}  & {{\scriptsize 2302.14}}  & {{\scriptsize 2301.90}} & {{\scriptsize 2288.82}}  &   {{\scriptsize 2296.38}}  & {{\scriptsize 2295.36}}
\\\hline
{\scriptsize$\Delta{\rm DIC}$}  & {-}  & {{\scriptsize -2.36}} & {{\scriptsize +10.88}}  &   {{\scriptsize +5.52}}  & {{\scriptsize +6.25}}
\\\hline
\end{tabular}}
\end{center}
\caption{Same as in Table 1, but also considering the prior on $H_0=(73.5\pm 1.4)$ km/s/Mpc from SH0ES \cite{Riess2019}.}
\label{tableFit1}
\end{table*}

Let us therefore consider a generic cosmological framework described by the spatially flat Friedmann-Lema\^\i tre-Robertson-Walker  (FLRW) metric. The vacuum energy density in the RVM can be written in the form \,\cite{JSPRev2013,SolGom2015}:
\begin{equation}\label{eq:RVMvacuumdadensity}
\rv(H) = \frac{3}{8\pi{G}_N}\left(c_{0} + \nu{H^2+\tilde{\nu}\dot{H}}\right)+{\cal O}(H^4)\,,
\end{equation}
in which the ${\cal O}(H^4)$ terms will be neglected for the physics of  the post-inflationary epoch.
The above generic structure can be motivated from the aforementioned explicit QFT calculations on a FLRW background\,\cite{Cristian2020}.
The additive constant $c_0$ is fixed by the boundary condition $\rho_{\rm vac}(H_0)=\rvo$.  Notice that the two dynamical components $H^2$ and $\dot{H}$ are dimensionally homogeneous and, in principle, independent.
Their associated  (dimensionless) coefficients $\nu$ and $\tilde{\nu}$ encode the dynamics of the vacuum at low energy  and we naturally expect $|\nu,\tilde{\nu}|\ll1$. An estimate of $\nu$ in QFT indicates that it is of order $10^{-3}$ at most \cite{Fossil07}. In the calculation of  \cite{Cristian2020} these coefficients are expected to be of order $\sim M_X^2/\mPl^2\ll 1$, where  $\mPl\simeq 1.22\times 10^{19}$ GeV is the Planck mass and $M_X$  is of order of a typical Grand Unified Theory (GUT) scale, times a multiplicity factor accounting for the number of heavy particles in the GUT.
We will be particularly interested in the RVM density obtained from the choice $\tilde{\nu}=\nu/2$.  As a result, $\rv(H) ={3}/(8\pi G_N)\left[c_0 + \nu\left({H^2+\frac12\dot{H}}\right)\right]$. We will call this form of the VED the  `RRVM' since it realizes the generic RVM density \eqref{eq:RVMvacuumdadensity} through the Ricci scalar $\mathcal{R} = 12H^2 + 6\dot{H}$, namely
\begin{equation}\label{RRVM}
\rv(H) =\frac{3}{8\pi{G_N}}\left(c_0 + \frac{\nu}{12}\mathcal{R}\right)\equiv \rv(\mathcal{R})\,.
\end{equation}
Such a RRVM implementation has the advantage that it gives a safe path to the early epochs of the cosmological evolution since in the radiation dominated era we have $\mathcal{R}/H^2\ll 1$, and hence we do not generate any conflict with the BBN nor with any other feature of the modern universe.  Of course, early on the RVM has its own mechanism for inflation (as we have already mentioned), but we shall not address these aspects here, see \cite{JSPRev2013,SolGom2015,LiBasSol2013, Sola2015,NickJoan2020}.

\subsection{Type I RRVM}

Friedmann's equation and the acceleration equation relate $H^2$ and $\dot{H}$ with the energy densities and pressures for the different species involved, and read
\begin{align}
3H^2 &= 8\pi{G_N}\left(\rho_m + \rho_{\rm{ncdm}} + \rho_\gamma + \rv(H)\right),\label{FriedmannEquation} \\
3H^2 + 2\dot{H} &= -8\pi{G_N}\left(p_{\rm{ncdm}} + p_\gamma + p_{\rm vac}(H)\right)\label{AccelerationEquation}\,.
\end{align}
The total nonrelativistic matter density is the sum of the cold dark matter (CDM) component and the baryonic one: $\rho_m = \rho_{dm} + \rho_b$.  The contributions of massive and massless neutrinos are included in $\rho_{\rm ncdm}$  (`${\rm ncdm}$' means non-CDM). Therefore the total (relativistic and nonrelativistic)  matter density is $\rho_t=\rho_m+ \rho_\gamma+\rho_{\rm ncdm}$. Similarly, the total matter pressure reads  $p_t=p_{\rm{ncdm}} + p_\gamma$  (with  $p_\gamma=(1/3)\rho_\gamma$).
We note that  there is a transfer of energy from the relativistic neutrinos to the nonrelativistic ones along the whole cosmic history, and hence it is not possible (in an accurate analysis) to make a clear-cut separation between the two.  Our procedure adapts to our own modified version of the system solver \texttt{CLASS}\,\cite{CLASS}.  The latter solves the coupled system of Einstein's and Boltzmann's differential equations  for any value of the scale factor and, in particular, provides the functions
$\rho_h = \rho_{\rm ncdm} - 3p_{\rm ncdm}$ and $\rho_\nu = 3p_{\rm ncdm}$ for the nonrelativistic and relativistic neutrinos, respectively.
This allows to compute the combination $\mathcal{R}/12=H^2 + (1/2)\dot{H}$  appearing in \eqref{RRVM}  in terms of the energy densities and pressures using \eqref{FriedmannEquation} and \eqref{AccelerationEquation}:
\begin{equation}\label{combination}
\mathcal{R} = 8\pi{G_N}\left(\rho_m + 4\rv + \rho_h\right)\,.
\end{equation}
Notice that the photon contribution cancels exactly in this expression and hence $\rv$ from \eqref{RRVM} remains much smaller than the photon density in the radiation epoch, entailing no alteration of the thermal history.  While neutrinos do not behave as pure radiation for the aforementioned reasons,
one can check numerically (using \texttt{CLASS}) that the ratio $r \equiv{\rho_h}/{\rho_m}$ is very small throughout the entire cosmic history up to our time (remaining always below $10^{-3}$). Thus, we can neglect it in \eqref{RRVM} and we can solve for the vacuum density as a function of the scale factor $a$ as follows:
\begin{equation}\label{eq:VacuumDens}
\rv(a) = \rvo + \frac{\nu}{4(1-\nu)}(\rho_m(a) - \rho^0_m)\,,
\end{equation}
where ` $0$' (used as subscript or superscript) always refers to current quantities.  For $a=1$ (today's universe)  we confirm the correct normalization:  $\rv(a=1) = \rvo$. Needless to say,  $\rho_m(a)$ is not just $\sim a^{-3}$ since the vacuum is exchanging energy with matter here.  This is obvious from the fact that the CDM exchanges energy with the vacuum (making it dynamical):
\begin{equation}\label{eq:LocalConsLaw}
\dot{\rho}_{dm} + 3H\rho_{dm} = -\dot{\rho}_{\rm vac}\,.
\end{equation}
Baryons do not interact with the vacuum, which implies $\dot{\rho}_b + 3H\rho_b =0$, and as a result the total matter contribution ($\rho_m$)  satisfies the same local conservation law \eqref{eq:LocalConsLaw} as CDM:
$\dot{\rho}_m + 3H\rho_m = -\dot{\rho}_{\rm vac}$.  Using it with \eqref{eq:VacuumDens} we find
$\dot{\rho}_{m} + 3H\xi\rho_m = 0$, where we have defined  $\xi \equiv \frac{1 -\nu}{1 - \frac{3}{4}\nu}$.
Since $\nu$ is small, it is convenient to encode the deviations with respect to the standard model in terms of the effective parameter $\nueff\equiv\nu/4$:
\begin{equation}
\xi = 1-\nu_{\rm eff} + \mathcal{O}\left(\nu_{\rm eff}^2\right)\,.
\end{equation}
It is straightforward to find the expression for the matter densities:
\begin{equation}\label{eq:MassDensities}
\rho_m(a) = \rho^0_m{a^{-3\xi}}\,, \ \ \ \rho_{dm}(a) = \rho^{0}_m{a^{-3\xi}}  - \rho^0_b{a^{-3}} \,.
\end{equation}
They recover the $\CC$CDM form for $\xi=1$ ($\nueff=0$).  The small departure is precisely what gives allowance for a mild dynamical vacuum evolution:
\begin{align}  \label{Vacdensity}
\rv(a) &= \rvo + \left(\frac{1}{\xi} -1\right)\rho^0_m\left(a^{-3\xi} -1\right)\,.
\end{align}
The vacuum  becomes rigid only for $\xi=1$ ($\nueff=0$).

\subsection{Type II RRVM}
For type II models matter is conserved (no exchange with vacuum), but the vacuum can still evolve provided the gravitational coupling also evolves (very mildly) with the expansion: $G=G(H)$. Following the notation of \cite{BD-RVM}, let us define (just for convenience) an auxiliary variable  $\varphi=G_N/G $ -- in the manner of  a Brans-Dicke {(BD)} field\,{\cite{BD}}, without being really so. Notice that $\varphi\neq 1$ in the cosmological domain, but remains very close to it, see Tables 1 and 2. 
{For convenience, in the last column of Tables 1 and 2 (and Fig.2) we include the updated results of \cite{BD-RVM} (BD model with a cosmological constant) with the data changes indicated in the caption of  Table 1.}

Friedman's equation for type-II model takes the form 
\begin{equation}\label{eq:fried}
3H^2=\frac{8\pi G_N}{\varphi}\left[\rho_t+C_0+\frac{3\nu}{16\pi G_N}(2H^2+\dot{H})\right]\,,
\end{equation}
with $C_0=3c_0/(8\pi G_N)$.
The Bianchi identity dictates the correlation between the dynamics of $\varphi$ and that of $\rv$ \footnote{{For type-II models the running of $G$ is triggered by that of $\rv$ via the Bianchi identity \eqref{eq:Bianchi}. If matter is self-conserved,
such running is unavoidable from the existence of the quantum effects  $\sim H^2$ (and/or $\dot{H}$) inducing the running of $\rv$, see \cite{Cristian2020}.
This does not exclude other microscopic mechanisms, but for type-II the running of $G$  is necessary to comply with general covariance. In the BD case  \cite{BD-RVM}, instead,
 $\varphi$ is an explicit field ingredient of the classical action.}}:
\begin{equation}\label{eq:Bianchi}
\frac{\dot{\varphi}}{\varphi}=\frac{\dot{\rho}_{\rm vac}}{\rho_t+\rv}\,,
\end{equation}
where $\rho_t $ is as before the total matter energy density and $\rv$ adopts exactly the same form as in \eqref{RRVM}. Using these equations one can show that the approximate behavior of the VED in the present time is (recall that $|\nueff|\ll1$):
\begin{equation}\label{eq:VDEm}
\rv(a)=C_0(1+4\nueff)+\nueff\rho_m^{0}a^{-3}+\mathcal{O}(\nueff^2)\,.
\end{equation}
Again, for $\nueff=0$ the VED is constant, but otherwise it shows a moderate dynamics of ${\cal O}(\nueff)$ as in the type I case \eqref{Vacdensity}.
Here, however, the exact solution must be found numerically.  One can also show that the behavior of $\rv(a)$ in the radiation dominated epoch is also of the form \eqref{eq:VDEm}, except that the constant additive term can be completely neglected.  It follows that  $\rv(a)\ll\rho_r(a)=\rho_r^0 a^{-4}$  for $a\ll 1$ and hence the VED for the type II model does not perturb the normal thermal history (as in the type I model). Finally, one finds $\varphi(a)\propto a^{-\epsilon}\approx 1-\epsilon\ln\,a$ in the current epoch (with $0<\epsilon\ll 1$ of order $\nueff$), thus confirming the very mild (logarithmic) evolution of $G$.
%
\begin{figure*}
\centering
\includegraphics[angle=0,width=0.9\linewidth]{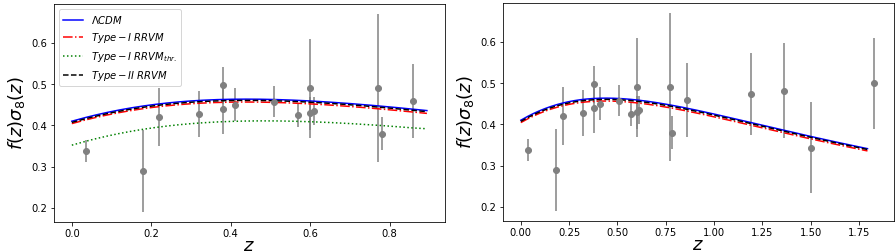}
\caption{\label{fig:fs8Evolution}%
Theoretical curves of $f(z)\sigma_8 (z)$ for the various models and the data points employed in our analysis, in two different redshift windows. To generate this plot we have used the central values of the cosmological parameters shown in Table 1. The type I running vacuum model with threshold redshift $\zstar\simeq 1$  has a most visible and favorable impact on solving the $\sigma_8$ tension.
}
\end{figure*}
%


\section{Threshold redshift scenario for type I models}

One possibility that has been explored in the literature in different type of models is to admit that the dynamics of vacuum is relatively recent (see e.g. \cite{Salvatelli2014}).  This means to study the consequences of keeping deactivated the interaction between the vacuum energy density and the CDM for most of cosmic history until the late universe when the DE becomes apparent. We denote the threshold value of the scale factor when the activation takes places by $\astar$. According to this scenario the VED was constant prior to $a=\astar$ and it just started to evolve for $a>\astar$.  While $\rv$ is a continuous function, its derivative is not since we mimic such situation through a Heaviside step function  $\Theta(a-\astar)$.  If we would have a microscopic description of the phenomenon it should not be necessary to assume such a sudden (finite) discontinuity. However, a  $\Theta$-function description will be enough for our purposes.  Therefore, we assume that in the range   $a<\astar$  (hence for $z> \zstar$) we have
%
\begin{align}
&\rho_{dm}(a) = \rho_{dm}(\astar)\left(\frac{a}{\astar}\right)^{-3},  \ \ \ \ \ \ \ \ \ \  \nonumber\\
&\rv (a) = \rv(\astar)=\text{const.} \ \ \ \ \ \ \ \ \ \ \ \ \ \   (a<\astar)\,, \label{Vacdensitystar}
\end{align}
where $\rho_{dm}(\astar)$ and  $\rv(\astar)$ are computed from \eqref{eq:MassDensities} and \eqref{Vacdensity}, respectively.
%
%
In the complementary range, instead,  i.e. for  $a>\astar$   ($0<z<\zstar$) near our time,  the original equations \eqref{eq:MassDensities} and \eqref{Vacdensity} hold good.

Notice that the above threshold procedure is motivated specially within type I models in order to preserve the canonical evolution law for the matter energy density when the redshift is sufficiently high. In fact, the threshold redshift value need not be very large and as we shall see in the next section,  if fixed by optimization it turns out to be of order  $\zstar\simeq 1$.  Above it ($z>\zstar$) the matter density evolves as in the $\CC$CDM and in addition $\rv$ remains constant. Its dynamics is only triggered at (and below) $\zstar$.  An important consequence of  such threshold is that the cosmological physics during the  CMB epoch (at $z\simeq 1000$) is exactly as in the $\CC$CDM.  For type II models there is still some evolution of the VED at the CMB epoch, but the matter density follows the same law as in the $\CC$CDM case. For this reason we will not investigate here the threshold scenario for type II models.

\section{Cosmological perturbations}   

So far so good for the background cosmological equations in the presence of dynamical vacuum. However,  an accurate description of the  large scale structure  (LSS) formation data is also of paramount importance, all the more if we take into account that one of the aforementioned $\CC$CDM tensions (the $\sigma_8$ one) stems from it.  Allowing for some evolution of the vacuum can be the clue to solve the $\sigma_8$ tension since such dynamics affects nontrivially the cosmological perturbations\,\cite{rvmPerturb}.
We consider the perturbed, spatially flat,  FLRW metric   $ds^2=-dt^2+(\delta_{ij}+h_{ij})dx^idx^j$, in which $h_{ij}$ stands for the metric fluctuations. These fluctuations are coupled to the  matter density perturbations $\delta_m=\delta\rho_m/\rho_m$.
We shall refrain from providing details of this rather technical part, which will be deferred for an expanded presentation  elsewhere.  However, the reader can check e.g.  \cite{rvmfit1,rvmPerturb,rvmfit2,GomSolBas2015,BD-RVM} for the basic discussion of the RVM perturbations equations. The difference is that here we have implemented the full perturbations analysis  in the context of the Einstein-Boltzmann code \texttt{CLASS}\,\cite{CLASS}  (in the synchronous gauge\,\cite{MaBertschinger}).   Let us nonetheless mention a few basic perturbations equations which have a more direct bearing on the actual fitting analysis presented in our tables and figures.  Since baryons do not interact with the time-evolving VED the perturbed conservation equations are not directly affected. However, the corresponding equation for CDM is modified in the following way:
\begin{equation}\label{eq:perturbCDM}
\dot{\delta}_{dm}+\frac{\dot{h}}{2}-\frac{\dot{\rho}_{\rm vac}}{\rho_{dm}}\delta_{dm}=0\,,
\end{equation}
with $h=h_{ii}$ denoting the trace of $h_{ij}$. We remark that the term $\dot{\rho}_{\rm vac}$ is nonvanishing for these models and affects the fluctuations of CDM in a way which obviously produces a departure from the $\CC$CDM. The above equation is, of course, coupled with the metric fluctuations and the combined system must be solved numerically.

The analysis of the linear LSS regime is performed with the help of the weighted linear growth $f(z)\sigma_8(z)$, where $f(z)$ is the growth factor and $\sigma_8(z)$ is the rms mass fluctuation amplitude on scales of $R_8=8\,h^{-1}$ Mpc at redshift $z$.   The quantity $\sigma_8(z)$ is directly provided by \texttt{CLASS} and the calculation of $f(a)$ (with $z=a^{-1}-1$ in our normalization) can be obtained as follows. If $\vec{k}$ denotes the comoving wave vector and $\vec{k}/a$  the physical one,  at  subhorizon scales its modulus (square) satisfies  $k^2/a^2\gg  H^2$. If, in addition, we are in the linear regime the matter density contrast can  be written as $\delta_m(a,\vec{k}) = D(a)F(\vec{k})$\,\cite{DodelsonBook,DEBook},
where the dependence on $\vec{k}$ factors out. The properties of $F(\vec{k})$ are determined by the initial conditions and $D(a)$ is called the growth function.  The relation between the matter power spectrum and the density contrast reads
$P_m(a,\vec{k}) = C\langle\delta_m(a,\vec{k})\delta^{*}_m(a,\vec{k})\rangle  \equiv D^2(a)P(\vec{k})$,
where $C$ is a constant and $P(\vec{k}) =C\langle{F(\vec{k})}F^{*}(\vec{k})\rangle$ is the primordial power spectrum (determined from the theory of inflation).  Since neither $F(\vec{k})$ nor  $P(\vec{k})$  depend on $a$, the linear growth $f(a) = d\ln \delta_m(a,\vec{k})/d\ln a$ is given by $f(a)={d\ln D(a)}/{d\ln a}$, and ultimately by
\begin{equation}
f(a) =\frac{d\ln P^{1/2}_m(a,\vec{k})}{d\ln a} = \frac{a}{2P_m(a,\vec{k})}\frac{dP_m(a,\vec{k})}{da}\,.
\end{equation}
It follows that we may extract the  (observationally measured) linear growth function  $f(a)$ directly from the matter power spectrum $P_m(a,\vec{k})$, which is computed numerically by \texttt{CLASS}  for all  values of $a$ and $\vec{k}$ (assuming   adiabatic initial conditions). This allows us to compare theory and observation for the important  LSS part.

%
\begin{figure*}
\centering
\includegraphics[angle=0,width=0.9\linewidth]{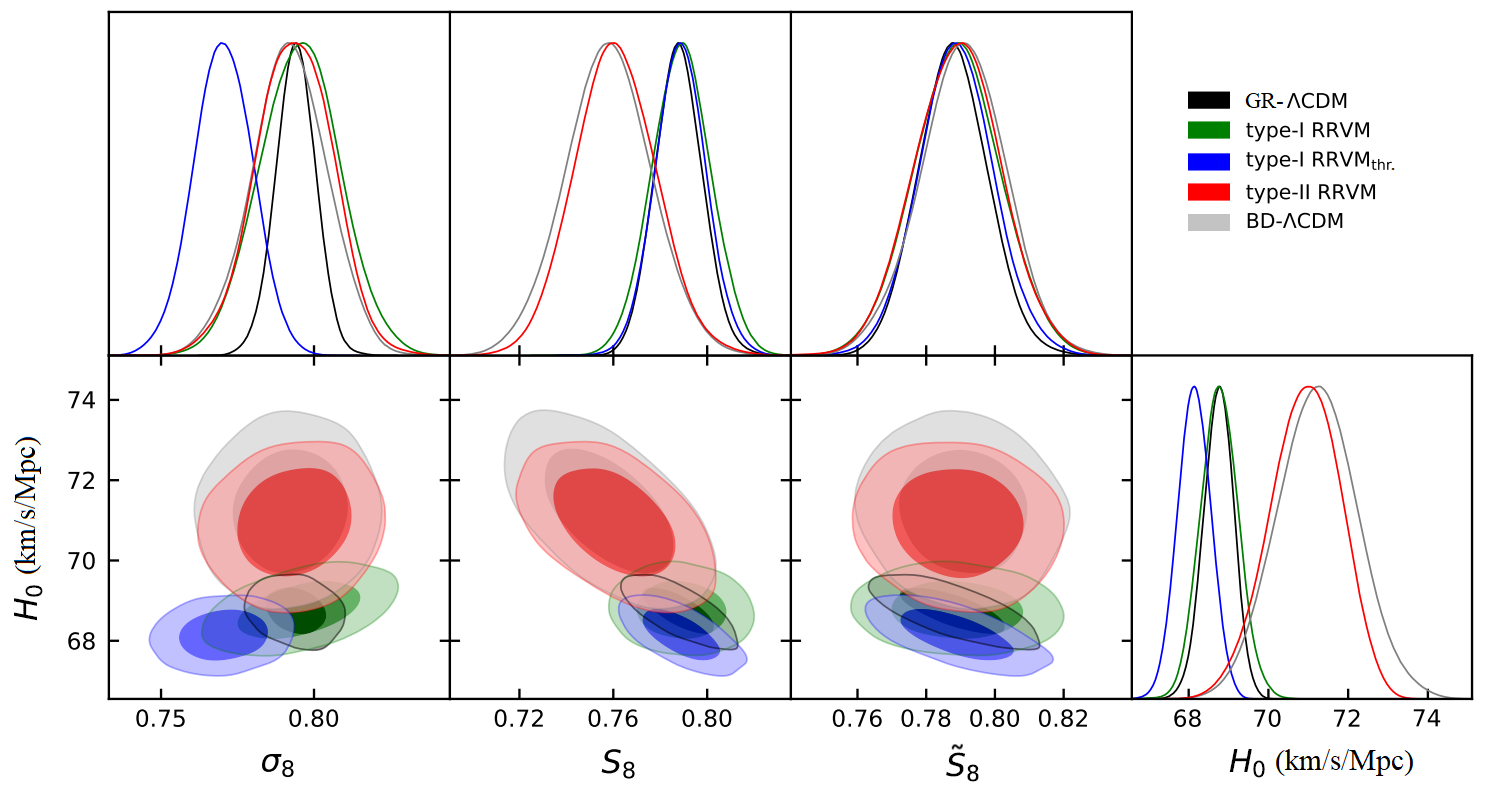}
\caption{\label{fig:XCDMEvolution}%
$1\sigma$ and $2\sigma$ contours in the $H_0$-$\sigma_8,S_8,\tilde{S}_8$ planes and the corresponding one-dimensional posteriors for the {GR- and BD- $\Lambda$CDM and the RRVM's} obtained from the fitting analyses with our Baseline+$H_0$ data set. The type II model manifestly alleviates the $H_0$ tension without spoiling the $\sigma_8$ one (even if phrased through the alternative parameters $S_8$ or $\tilde{S}_8$, see text), whereas the type I model with threshold redshift $\zstar\simeq 1$ can fully solve the latter (see also Fig.\,1) but cannot address the former.
}
\end{figure*}
%

\section{Fitting results and discussion}

To compare the  RRVM's (types I and II) with the $\CC$CDM, we have defined a joint likelihood function ${\cal L}$. The overall fitting results are reported in Tables 1 and 2. The used data sets are the same as those described  in detail in  Ref. \cite{BD-RVM}, except the updated values pointed out in the caption of Table 1.  Assuming Gaussian errors, the total $\chi^2$ to be minimized in our case is given by
\be
\chi^2_{\rm tot}=\chi^2_{\rm SNIa}+\chi^2_{\rm BAO}+\chi^2_{ H}+\chi^2_{\rm f\sigma_8}+\chi^2_{\rm CMB}\,.
\ee
The above $\chi^2$ terms are defined in the standard way from the data including the covariance matrices\,\cite{DEBook}.  In particular, the $\chi^2_{H}$ part may contain or not the local $H_0$ value measured by Riess et al.\,\cite{Riess2019} depending on the setup indicated in the tables (apart from the cosmic chronometer data employed also in \cite{BD-RVM}). The local determination of $H_0$ (which is around $4\sigma$ away from the corresponding Planck 2018 value based on the CMB) is the origin of the so-called $H_0$ tension\,\cite{tensions,Intertwined}.
Taking into account that the RRVM's of type I and II have one and two more parameters, respectively, as compared to the $\CC$CDM, a fairer  model comparison is achieved by computing the differences between the Deviance Information Criterion \cite{DIC}, of the $\CC$CDM model and the RRVM's: $\Delta{\rm DIC}={\rm DIC}_{\rm \CC CDM}-{\rm DIC}_{\rm RRVM}$.  These differences will be (and in fact are) positive if the RRVM's fit better the overall data than the $\CC$CDM. The DIC is defined as
\begin{equation}
{\rm DIC}=\chi^2(\overline{\theta})+2p_D\,.
\end{equation}
Here $p_D=\overline{\chi^2}-\chi^2(\overline{\theta})$ is the effective number of parameters of the model, and $\overline{\chi^2}$ and $\overline{\theta}$ the mean of the overall $\chi^2$ distribution and the parameters, respectively. The DIC is a  good approximation to the exact Bayesian approach and works optimal if the posterior distributions are sufficiently Gaussian.  To obtain the posterior distributions and corresponding constraints for the various dataset combinations we have used  the Monte Carlo cosmological parameter inference code \texttt{Montepython}\cite{Montepython} in combination with the mentioned Einstein-Boltzmann code \texttt{CLASS}\,\cite{CLASS}.

The value of DIC can be computed directly from the Markov chains generated with \texttt{MontePython}.
For values $+5<\Delta{\rm DIC} <+10$ we would conclude strong evidence of the RRVM's as compared to the $\CC$CDM, and for $\Delta{\rm DIC}>+10$ the evidence is very strong. Such is the case when we use a threshold  redshift $\zstar\simeq 1$ in type I RRVM (cf. Tables 1 and 2). In contrast, when the threshold is removed we find only moderate evidence against it ($-3<\Delta{\rm DIC} <-2$), although the fitting performance keeps on being slightly better (smaller $\chi^2_{\rm min}$) than the GR-$\Lambda$CDM, similar to e.g. coupled dark energy \cite{PRD2020}. Quite obviously, the effect of the threshold can be very important and indicates that a mild dynamics of the vacuum is very much welcome, especially if it is activated at around the very epoch when the vacuum dominance appears, namely at around $z\simeq 1$. To be more precise, the vacuum dominance in the $\CC$CDM starts at around $z\simeq 0.3$.  Therefore, these results suggest that if the vacuum starts to be slightly dynamical at an earlier point  which is `close' (in redshift terms) to the transition from deceleration to acceleration ($z\simeq 0.7$), then the impact on the description of the overall SNIa+BAO+$H(z)$+LSS+CMB data becomes extraordinarily significant on statistical terms. Before the transition point, physics can remain basically unaltered with respect to the standard $\Lambda$CDM model, but the vacuum dynamics allows to suppress an exceeding amount of LSS in the universe, leading to a better description of the $f(z)\sigma_8(z)$ data set. It is not just that the total  $\chi^2_{\rm min}$ is 13 to 18 units smaller as compared to the $\CC$CDM in the presence of the threshold $\zstar$ (cf. Tables 1 and 2), but the fact that the information criteria (which take into account the penalty to be paid by the RRVM's for having more parameters) still decides very strongly in its favor. In the absence of the $H_0$ prior \cite{Riess2019}, type II RRVM performs a bit better than the GR-$\Lambda$CDM (cf. Table 1), but the improvement is not sufficient. Occam's razor penalizes the model for having two additional parameters than GR-$\Lambda$CDM and leads to a moderately negative evidence against it. When we include the prior, however, we get a strong evidence in its favor ($\Delta {\rm DIC}\gtrsim +5$, cf.\,Table 2), since this model can accommodate higher values of the Hubble parameter and hence loosen the $H_0$ tension. This is similar to what we found in \cite{BD-RVM} for Brans-Dicke cosmology with $\CC\neq 0$.

Finally, we want to remark a few things about the RRVM's under study, in connection with the cosmological tensions, cf. Tables 1 and 2, and the contours in Fig. 2: (i) the only model capable of alleviating the $H_0$ tension is RRVM of type II; (ii) the values of $S_8$ in all RRVM's are perfectly compatible with recent weak lensing and galaxy clustering measurements \cite{Heymans2020}. For type II a related observable analogous to (but different from) $S_8$ is possible: $\tilde{S}_8\equiv S_8/\sqrt{\varphi_0}$. It is connected with the time variation of $G=G_N/\varphi$ and can be viewed also as a rescaling $\Omega_m^0\to \Omega_m^0/\varphi_0$ in the effective Friedmann's equation for type II models, see Ref.\,\cite{BD-RVM}. We show the corresponding countours in Fig. 2; (iii) Quite remarkable is the fact that the value of $\sigma_8$ is significantly lower in the type I RRVM$_{\rm thr.}$ to the point that the $\sigma_8$ tension can be fully accounted for. We have checked that this feature is shared by the more general RVM class \eqref{eq:RVMvacuumdadensity} using the same threshold redshift.


\section{Conclusions}

We find significant evidence that a  mild dynamics of the cosmic vacuum would be helpful to describe the overall cosmological observations as compared to the standard cosmological model with  a  rigid $\CC$-term.  For type I models the level of evidence is very strongly supported by the DIC criterion provided there exists a threshold  redshift $\zstar\simeq 1$ where the vacuum dynamics is triggered. With such dynamics the $\sigma_8$ tension is rendered virtually nonexistent ($\lesssim 0.4\sigma$) \cite{Heymans2020}. The $H_0$ tension, however,  can only be improved within the type II model  with variable $G$ {and also with the related BD-$\CC$CDM model\cite{BD-RVM}}. For both the two tensions can be dealt with at a  time, the $H_0$ remaining at $\sim 1.6\sigma$ \cite{Riess2019} and the $\sigma_8$  one at $\sim 1.3\sigma$ (or at only $\sim 0.4\sigma$ if stated in terms of $S_8$)  \cite{Heymans2020}. The simultaneous alleviation of the two tensions is remarkable and is highly supported by the DIC criterion.

\acknowledgments

\noindent  JSP, JdCP and CMP are partially supported by  MINECO (Spain), SGR (Generalitat de Catalunya) and MDM (ICCUB).  AGV is funded by DFG (Germany).  JdCP and CMP are also supported by  FPI and FI fellowships, respectively.   JSP   also acknowledges the COST Association Action QG-MM. We thank H. Gil-Mar\'in for discussions.

\end{document}